# Van der Waals epitaxy of $Bi_2Se_3$ on Si(111) vicinal surface: An approach to prepare high-quality thin films of topological insulator

H.D. Li[1], Z.Y. Wang[1], X. Kan[1], X. Guo[1], H.T. He[2], Z. Wang[2], J.N. Wang[2], T.L. Wong[2], N. Wang[2], and M.H. Xie[1]

[1]*Physics Department, The University of Hong Kong, Pokfulam Road, Hong Kong, China*

[2]*Physics Department, Hong Kong University of Science and Technology, Clear Water bay, Kowloon, Hong Kong, China*

*Abstract*

Epitaxial growth of topological insulator $Bi_2Se_3$ thin films on nominally flat and vicinal Si(111) substrates is studied. In order to achieve planner growth front and better quality epifilms, a two-step growth method is adopted for the van der Waal epitaxy of $Bi_2Se_3$ to proceed. By employing vicinal Si(111) substrate surfaces, the in-pane growth rate anisotropy of $Bi_2Se_3$ is explored to achieve single crystalline $Bi_2Se_3$ epifilms, in which threading defects and twins are effectively suppressed. Optimization of the growth parameters has resulted in vicinal $Bi_2Se_3$ films showing a carrier mobility of ~ 2000 $cm^2V^{-1}s^{-1}$ and the background doping of ~ $3 \times 10^{18}$ $cm^{-3}$ of the as-grown layers. Such samples not only show relatively high magnetoresistance but also a linear dependence on magnetic field.

Author contributions: HDL, ZYW, XK, and XG conducted the MBE growth, RHEED, STM, and XRD experiments; HTH, ZW and JNW did the Hall and magneto-transport measurements, TLW and NW carried out the TEM study. MHX coordinated the project and is the corresponding author. His E-mail address is: mhxie@hkusua.hku.hk

*Main Text*

$Bi_2Q_3$ (Q = Se, Te) compounds are classical thermoelectric materials with high figure-of-merit, ZT (e.g., ZT ≈ 1 for $Bi_2Te_3$).*[1-3]* Recently, a huge resurgence of interests towards these materials is witnessed after a theoretical prediction and experimental revelation that such compounds are superior three-dimensional (3D) topological insulators (TIs).*[4, 5]* The crystals of $Bi_2Q_3$ have simple surface states in their bulk energy gaps, which are topologically protected. The spin states of surface electrons are coupled with momentum, which would have great potentials for future spintronic applications. As the bulk energy gap of $Bi_2Se_3$ is as large as ~ 0.3 eV, the crystal of $Bi_2Se_3$ is of particular interest for room temperature applications.

As-grown bulk crystals of $Bi_2Se_3$ usually show high concentrations of electrical carriers (electrons) and the Fermi levels are pinned in the conduction band.*[5, 6]* This hinders easy access of the TI states of the material. One may reduce the level of background doping by introducing acceptors in crystal,*[7, 8]* however, such an approach is not appropriate when the background carrier density is high. A high density of dopants introduces scattering centers, which degrade the transport properties of the material. For quantum device applications, it is desirable to fabricate thin films or nanostructures of $Bi_2Se_3$. However, the situation appears similar to bulk crystals in that the background electron concentrations higher than ~ $10^{19}$ $cm^{-3}$ are consistently observed.*[9-13]* It becomes imperative to obtain high quality films with low background doping for some practical purposes. In this study, we adopt the method of molecular-beam epitaxy (MBE) operating in ultrahigh vacuum to grow $Bi_2Se_3$ layers on Si(111) substrate. Such a method has recently been demonstrated to successfully grow thin films of $Bi_2Se_3$ *[14]* and $Bi_2Te_3$.*[15]* Here, we explore the method of van der Waal epitaxy and report on the effectiveness of using a vicinal surface to achieving high quality epitaxial layers.

The rhombohedral crystal of $Bi_2Se_3$ has the layered structure along the [111] crystallographic direction, where five atomic layers in the sequence of - Se - Bi - Se - Bi - Se - form a quintuple layer (QL) unit. Atoms within the QL unit are chemically bonded, while between adjacent QLs, they are bonded by the weak van der Waal (vdW) force.*[16]*

Such a unique structure of $Bi_2Se_3$ would favor a two-dimensional (2D) growth along [111] by using the so-called van der Waal epitaxy (vdWe) method.[17-20] An essence of the vdWe is to prepare a substrate surface that does not contain dangling bonds. For Si(111)-(7 ×7) that we use in this study, surface dangling bonds do exist. We may saturate such dangling bonds by hydrogen atoms before $Bi_2Se_3$ deposition in order to achieve the vdWe mode,[21] but here we introduce another approach by which the surface effect of the substrate can be suppressed using a two-step procedure. Initial deposition of a thin amorphous seed layer of $Bi_2Se_3$ at low temperature (LT) is followed by high-temperature (HT) growth, and the later stage HT deposition on the Se-terminated $Bi_2Se_3$ seed layer proceeds via the vdWe mode. We will show that such a two-step approach has resulted in films that are superior to those grown using a single step process. To minimize in-plane twinning of the film, we employ vicinal substrate surface (denoted as v-Si(111) hereafter). By optimizing the growth parameters (temperature and flux), $Bi_2Se_3$ thin films with low temperature resistivity of ~ 1 mΩ cm, carrier mobility of ~ 2000 $cm^2V^{-1}s^{-1}$ and carrier concentration ~ $3 \times 10^{18}$ $cm^{-3}$ are achieved.

Prior to the LT deposition of $Bi_2Se_3$, Si(111)-(7 ×7) substrate surface is exposed to a flux of Bi, achieving the β-phase ($\sqrt{3} \times \sqrt{3}$) structure of the substrate.[22] Then $Bi_2Se_3$ deposition is initiated at ~ 100 K using a high Se/Bi flux ratio of Se : Bi = 10 : 1. The growth front is monitored by reflection high-energy electron diffraction (RHEED). Figure 1(a_i) shows the evolution of the RHEED intensity $I$ and the reciprocal lattice parameter $D$ as the deposition proceeded. The RHEED patterns taken along [1$\bar{1}$0] (The Miller indices are based on cubic Si throughout) at different growth stages are also given in Figure 1(a_ii) – 1(a_iv). As is seen, upon $Bi_2Se_3$ deposition at LT, a film with a lattice constant the same as that of $Bi_2Se_3$ is immediately formed. After 1 ~ 2 QLs growth, the RHEED pattern disappears, indicating a fully disordered or amorphous film. The RHEED intensity correspondingly drops and the lattice parameter becomes immeasurable. We let the growth to continue for a further 2~3 QLs before gradually increasing the temperature to 520 K, at which diffraction pattern characteristic of $Bi_2Se_3$ crystal re-appears suddenly at a total deposition coverage of ~ 7 QLs. The reciprocal lattice parameter ($D$) also recovers instantaneously to that of $Bi_2Se_3$, suggesting a crystal $Bi_2Se_3$ layer formation.

From the RHEED pattern of Figure 1(a_ii), however, we note the initial crystalline Bi$_2$Se$_3$ is textured, containing in-plane rotation domains in all directions. We then slightly lower the substrate temperature to 450 K for continuous growth of Bi$_2$Se$_3$. At this stage, the RHEED intensity starts to oscillate, suggesting the layer-by-layer growth mode. By comparing the oscillation period with film thickness, we establish that one period of the RHEED oscillation corresponds to one QL deposition. Notably, after 2 QLs growth, the RHEED pattern becomes very clean, characteristic of a single crystalline Bi$_2$Se$_3$. Although a weak 3D feature is also noted in the RHEED at this oscillation stage, the 3D feature gradually disappears as the growth continues. The improvement of the surface quality with film thickness is apparent from the RHEED evolution. The diffraction streaks become increasingly sharper while the background intensity weakens. Cross-sectional transmission electron microscopy study of a sample prepared by such two-step method [Figure 1(c)] shows the initial amorphous seed layer to retain during the subsequent HT growth stage. Therefore, the crystalline Bi$_2$Se$_3$ formed at later HT growth stage is by a self-crystallization or organization process, without much influence from the substrate. The growth proceeds via van der Waals epitaxy. Figure 2(a) shows a large area scanning tunneling microscopy (STM) image of the surface of a grown Bi$_2$Se$_3$. The surface is seen to consist of triangular step-and-terrace features, underlying an anisotropic lateral growth rates in different azimuthal directions. With reference to the lattices of the substrate, we establish that it grows at a faster rate in the <$\bar{1}\bar{1}2$> direction than along <11$\bar{2}$> direction. The steps on surface are 9.5 Å high, corresponding to one QL of Bi$_2$Se$_3$. On the other hand, there are oppositely orientated triangles with approximately equal proportion on surface, indicating the film to contain twin defects. Twin boundaries are inevitable, one of which is revealed in the zoom-in STM image as shown in the inset of Figure 2(a). Moreover, from STM measurements, we also observe step-terminations and spiral mounds that are associated with threading screw dislocations. One such a dislocation is indicated by the dashed arrow in Figure 2(a) inset, and the density of threading screws is estimated to be of the order of ~ $10^9$ cm$^{-2}$.

Despite all the defects, such a film grown by the two-step method is still superior to the ones grown using a single step method. If Bi$_2$Se$_3$ is grown continuously at LT, disordered or amorphous-like film is obtained. Post-growth annealing may lead to crystallization of such films, but it can hardly result in high quality crystals. On the other hand, if $Bi_2Se_3$ is grown at HT continuously from the beginning, layer-by-layer growth is indeed observable as judged from the RHEED intensity oscillations, but it is usually accompanied by some 3D features on the growth front. The surface smoothness, domain size and the crystallinity are inferior to that grown using the two-step approach. For example, Figure 1(a_v) shows a RHEED pattern of the HT-grown $Bi_2Se_3$ surface, revealing a spotty and split diffraction patterns characteristic of single-step-grown surfaces.

The existence of twins even in films grown by the two-step method is still undesirable. Noting that the lateral growth rate is anisotropic, we resort to vicinal substrate to bias the growth of one domain over the other.[23, 24] Using the same two-step approach but on a v-Si(111), the evolution of the RHEED intensity, reciprocal lattice parameter, and the RHEED patterns are given Figure 1(b). For the LT seed layer growth, a similar observation to that of the flat film growth is made. However, upon increasing the temperature for subsequent crystalline film formation, no RHEED intensity oscillation is recorded, suggesting the step-flow growth mode, possibly related to the presence of a high density of steps on surface due to the vicinal substrate. Unlike that on flat substrate, no 3D feature is ever noted during the HT growth stage on v-Si(111). The sharpness of the RHEED streaks and also the observation of Kikuchi lines indicate the film is of high structural quality.

The advantage of using v-Si(111) is more apparent when examining the morphology of the grown films. Figure 2(b) shows a STM image of such a surface, which displays a unique pointed-step structure. The pointed, rather than straight, step edges conform to the fast growth rate along $<\bar{1}\bar{1}2>$. The oppositely oriented twin domains would show relatively straight descending steps, while the ascending steps on the opposite side, if present, would be similarly pointed. Examples of the latter twin domain areas in Figure 2(b) are marked by the dashed triangles pointing to the left. The larger triangles pointing to the right mark the majority domains in sample. As is seen, the two domains of the twin defects are no longer equal in proportion and the left-pointing triangular domains are effectively suppressed by using the vicinal substrate. This may be understood by that lateral growth of such domains would be in the direction of ascending steps, which is arrested by step flows during material deposition. In the mean time, threading screw dislocations, observable only near the twin boundaries as exemplified in the inset of Figure 2(b), are diminished by using the vicinal substrate. An estimate of such threading defects from the STM measurements results in a value of ~ $10^8$ cm$^2$, which is an order of magnitude reduction over that in flat films. To characterize the phase purity and crystal quality of such vicinal films, X-ray diffraction (XRD) rocking curves are measured, an example of which is shown in Figure 2(c). A peak at ~ 12.71 ° is observed, corresponding nicely to the (0 0 6) diffraction of a Bi$_2$Se$_3$ crystal by noting the ~ 3.5 ° offcut angle of the substrate with respect to (111) plane (the theoretical Bragg angle is 9.32 °). This observation shows that there is little tilting of the epifilm from that of the substrate surface. The reflective θ-2θ scans of the sample shown in the inset of Figure 2(c) reveal diffraction peaks which are all index-able to a crystalline Bi$_2$Se$_3$.

The improvement of the surface and structural quality of Bi$_2$Se$_3$ by using vicinal substrate surfaces has also led to an improvement in the electronic properties of the films. Figure 3(a) compares the temperature dependence of film's resistivity between a flat (i) and a vicinal film (ii). The two samples were grown at exactly the same MBE condition (T = 450 K, Se : Bi = 10 : 1) and using the same two-step growth procedure. Data measured at temperatures above 150 K contain contributions from the substrate, so they are not shown here. It is shown that the resistivity of Bi$_2$Se$_3$ on v-Si(111) is consistently higher than that on flat substrate. Figure 3(b) shows temperature dependent carrier densities and mobility in the flat (i) and vicinal film (ii), respectively. A background electron concentration of ~ $3 \times 10^{19}$ cm$^{-3}$ is observed in the flat film, and the mobility is ~ 1200 cm$^2$V$^{-1}$s$^{-1}$ at 2 K. On the other hand, the vicinal film shows a carrier concentration of ~ $5 \times 10^{18}$ cm$^{-3}$ and a mobility of ~1600 cm$^2$V$^{-1}$s$^{-1}$ at the same temperature. The best sample we have measured shows a background electron density of $3 \times 10^{18}$ cm$^{-3}$ and a mobility of 2000 cm$^2$V$^{-1}$s$^{-1}$. Such a background carrier density represents more than one order of magnitude reduction over the best flat samples. As point defects are more likely related to growth conditions of the MBE [refer to Figures 3(c) and 3(d)], such a reduction of background doping in vicinal films is likely linked to the reduction of extended defects, such as twin boundaries and dislocations. The temperature-dependent Hall measurements (Figure 3(b)) reveal that the background carrier densities do not change much with temperature, indicating full ionization of the electron donors in $Bi_2Se_3$ even at LT. On the other hand, increasing the temperature has caused a notable decrease of electron mobility, implying phonon scattering processes. Finally, the MBE condition will affect point defect formation, causing additional modification of the electronic behavior of the films. Figures 3(c) and 3(d) show, respectively, the effects of both substrate temperature and Se-to-Bi flux ratio on the density and mobility of electrons in films. The samples appear superior when grown at temperatures in the range of 420 - 470 K (i.e., 1/5 $T_m$ - 1/3.5 $T_m$, where $T_m$ ~ 710 °C is the melting point of $Bi_2Se_3$) and at the flux ratio of 7 : 1 ~ 15 : 1. It is noted that under such growth conditions, the films also show good morphologies and structural quality. At too low a temperature, the crystallinity of the film degrades as indicated by a high density of dislocations and twin boundaries; while growth at too high a temperature causes the surface to contain cracks and holes, probably due to material decomposition. Although higher Se-to-Bi flux ratios (> 20 : 1) may consistently produce good surfaces of the epifilms, the electronic property degrades, probably due to creation of anti-site or interstitial Se in film. At low flux ratios (< 5 : 1), other more Bi-rich phase(s) of Bi-Se may form in addition to $Bi_2Se_3$, as revealed by some preliminary XRD measurements.

Finally, magneto-transport properties of the flat and vicinal $Bi_2Se_3$ films are studied. Fig. 4 (a) shows the normalized magnetoresistance (MR) $R/R(B = 0)$ of the flat (red circle) and vicinal (black square) $Bi_2Se_3$ films versus perpendicular magnetic field ($B$) up to 7 T at $T = 3$ K, respectively. One can see that the MR of both films at lower $B$ field can be well fitted by a parabolic dependence: $R/R(B = 0) = 1 + (\alpha * B)^2$, as shown by the solid fitting curves in Fig. 4(a). This dependence is believed to result from the Lorentz force deflection of carriers and $\alpha$ is proportional to the film mobility.[25] In addition, the vicinal film shows a much larger magnetoresistance than the flat one. In comparison with the flat film, the vicinal film has a larger mobility (see Fig. 3(b)) leading to larger $\alpha$ of parabolic MR observed. However, a crossover from this parabolic dependence to a linear dependence of MR is observed at higher *B* field. To see this crossover more clearly, Fig. 4(b) shows the MR of the vicinal film with the *B* field up to 14 T. The linear dependence of MR in higher *B* field is clearly observed at all temperatures from 3K to 160K. According to the theory by Abrikosov,*[26]* a gapless semiconductor with linear dependence of energy on crystal momentum would exhibit a linear MR in the quantum limit. This so-called "quantum linear magnetoresistance" (QLMR) predicts that the linear slope is independent of temperatures and proportional to $1/n_e^2$. Indeed, *R(B)* curves in Fig. 4(b) show almost identical linear slopes with the temperature ranging from 3K to 160 K. Furthermore, the vicinal film shows a larger linear slope as shown in Fig. 4(a), which is consistent with its lower carrier densities than that of the flat film. Therefore, we attempt to attribute the observed high field linear MR to this QLMR, which could arise from the Dirac electrons of the topologically-protected 2D surface states. *[4, 5]* The observation of QLMR as an indication of the existence of 2D topological surface states is also noted in a recent study of $Bi_2Se_3$ nanoribbons.*[27]*

In summary, to minimize the effect of Si substrate lattice, a two-step growth procedure is adopted, where the low-temperature amorphous seed layer of a few QLs thick is kinetically stabilized on Si(111), while the amorphous seed layer serves as the "substrate" of subsequent vdWe of crystalline $Bi_2Se_3$ at high temperature. By employing vicinal substrates, in-plane twinning of the epilayer is effectively suppressed due to directional flow of surface steps. By optimizing the growth conditions of MBE, crystalline $Bi_2Se_3$ thin films with LT resistivity of ~ 1 mΩcm, carrier mobility of ~ 2000 $cm^2V^{-1}s^{-1}$ and electron density of ~ $3 \times 10^{18}$ $cm^{-3}$ is achieved. The epitaxial $Bi_2Se_3$ film shows high magneto-resistance and its linear dependence on magnetic field, which are likely related to the topological surface states of the sample.

*Experimental*

Sample preparation and surface morphology measurements by STM were performed in a multi-chamber ultrahigh vacuum system, where the base pressure in the vacuum chambers was below $2 \times 10^{-10}$ mbar. The atomic fluxes of Bi and Se were provided from standard K-cells, which were calibrated by an ion-gauge based beam-flux-monitor (BFM). Nominal flat Si(111) and vicinal (3.5° offcut from (111) towards [$\bar{1}\bar{1}2$] direction) substrates were chemically cleaned and degassed overnight in vacuum before deoxidization at 1400 K to obtained the (7 × 7) reconstruction. Afterwards, the surface was exposed to a flux of Bi until the Bi-induced β-phase ($\sqrt{3} \times \sqrt{3}$) structure was obtained.*[22]* $Bi_2Se_3$ deposition was then initiated at ~100 K as cooled by liquid nitrogen. After 1 ~ 2 QLs deposition, the temperature of the substrate was increased by radiation from a set of tungsten filaments. The temperature was monitored by a thermocouple near the sample stage which had been calibrated a priori by the melting points of indium (443 K), selenium (493 K), and bismuth (543 K). The RHEED patterns and the evolution of its intensity, inter-streaks spacing (D) were recorded by a charge-coupled device camera. For room-temperature STM measurements, the tunneling current was 0.2 nA and the sample bias was +0.8 V throughout. *Ex situ* XRD measurements were conducted on a high resolution diffractometer from Bruker Inc., using the Cu-K$\alpha$1 X-ray source. The temperature-dependent Hall and magneto-resistance measurements were done in a Quantum Design PPMS system using the Hall bar devices. The Hall bars were fabricated by the standard photolithography technique with Cr(10nm)/Au(150nm) metal contacts deposited by thermal evaporation. Cross-section TEM examination was carried out in a JEOL2010F high-resolution transmission electron microscope working at 200kV, and the TEM specimen was prepared by a standard procedure of mechanical thinning following by Ar ion milling. .

*Acknowledgement*

The authors are grateful to W. K. Ho for technical support in the growth experiments, and to Stephen S. Y. Chui for the XRD experiment. The PPMS facilities used for magneto transport measurements, is supported by the Special Equipment Grant from the University Grants Committee of the Hong Kong Special Administrative Region (SEG_CUHK06). This work is financially supported partially by a Collaborative Research Fund of the Research Grant Council of Hong Kong Special Administrative Region, China, under the grant No. HKU 10/CRF/08, and a Seed Fund for Basic Research of HKU.

*Figure Captions:*

Figure 1. Evolution of the RHEED during $Bi_2Se_3$ deposition on nominal flat (a) and vicinal (b) Si(111) surfaces using the two-step procedure. Panel (i) presents the evolution of the RHEED intensity (*I*, red curve) and the spacing (*D*, blue curve) between two integer diffraction streaks as a function of deposition time or coverage. Panels (ii), (iii) and (iv) in (a) and (ii), (iii) in (b) show the corresponding RHEED patterns from the growing surfaces at different growth stages. Panel (v) in (a) shows the RHEED pattern of a surface prepared using a single step growth procedure. In (ii), the dashed arrows point to diffraction features originated from in-plane texture of the film, which are observable after annealing low-temperature-grown seed layers. The dashed squares in (iii) mark the location from where the RHEED intensity *I* is measured [Note: this location corresponds to an off-specular position but to a transmission diffraction spot of the bulk crystal, refer to panel (a_v)]. In (a_i) and (b_i), two horizontal dotted lines show the expected *D* values of Si and $Bi_2Se_3$ crystals. The RHEED patterns shown in (ii) have been logarithmly treated to enhance the weak diffraction features. (c) Cross-section TEM micrograph showing the amorphous seed layer between Si substrate and epitaxial $Bi_2Se_3$. The dashed lines mark the gaps between adjacent QLs of $Bi_2Se_3$ and the inset shows a schematic drawing of the layered structure of $Bi_2Se_3$ crystal.

Figure 2. Large area STM images (size: $2.5 \times 2.5$ μm$^2$) of $Bi_2Se_3$ films grown on (a) flat and (b) vicinal Si(111) substrates, respectively. The inset in (a) is a close-up image (size: $400 \times 400$ nm$^2$) of the flat surface showing the boundary (pointed by solid arrows) between two 180° twins, and a spiral mound (pointed by the dashed arrow) at a threading screw dislocation. In (b), the dashed triangles pointing to the right or left mark the oppositely oriented twin domains in the vicinal film. The inset in (b) is zoom-in image (size: $600 \times 600$ nm$^2$) of the vicinal surface showing threading dislocations near the domain boundaries (dashed arrows). The hexagons drawn at the bottom right of the main images indicate the crystallographic direction of the surface lattice, and it is along [$\bar{1}\bar{1}2$] from left to right. (c) XRD rocking curve of a vicinal $Bi_2Se_3$ film and the inset show the reflective θ-2θ scan of the same sample.

Figure 3. (a) Resistivity ($\rho_{xx}$) as a function of temperature of the as-grown $Bi_2Se_3$ film on (i) flat and (ii) vicinal Si(111) surfaces, respectively. (b) Hall density ($n_e$) and mobility ($\mu_e$) of electrons in (i) flat and (ii) vicinal films. (c) Dependence of background electron density ($n_e$, red squares) and mobility ($\mu_e$, blue circles) on deposition temperature, and (d) on Se/Bi flux ratio of MBE. Data in (c) and (d) are obtained by Hall measurement at ~100 K.

Figure 4. (a) The normalized resistance $R/R(B=0)$ of flat and vicinal films as a function of perpendicular magnetic field at $T = 3K$. At low $B$ filed, $R(B)$ curve shows a $B^2$ dependence, as shown by the solid fitting lines. (b) $R(B)$ curves measured in the vicinal film with temperatures ranging from 3K to 160 K and $B$ field up to 14 T. Linear $B$ dependence at high magnetic field is clearly seen at each temperature.

*Figures*

Figure 1

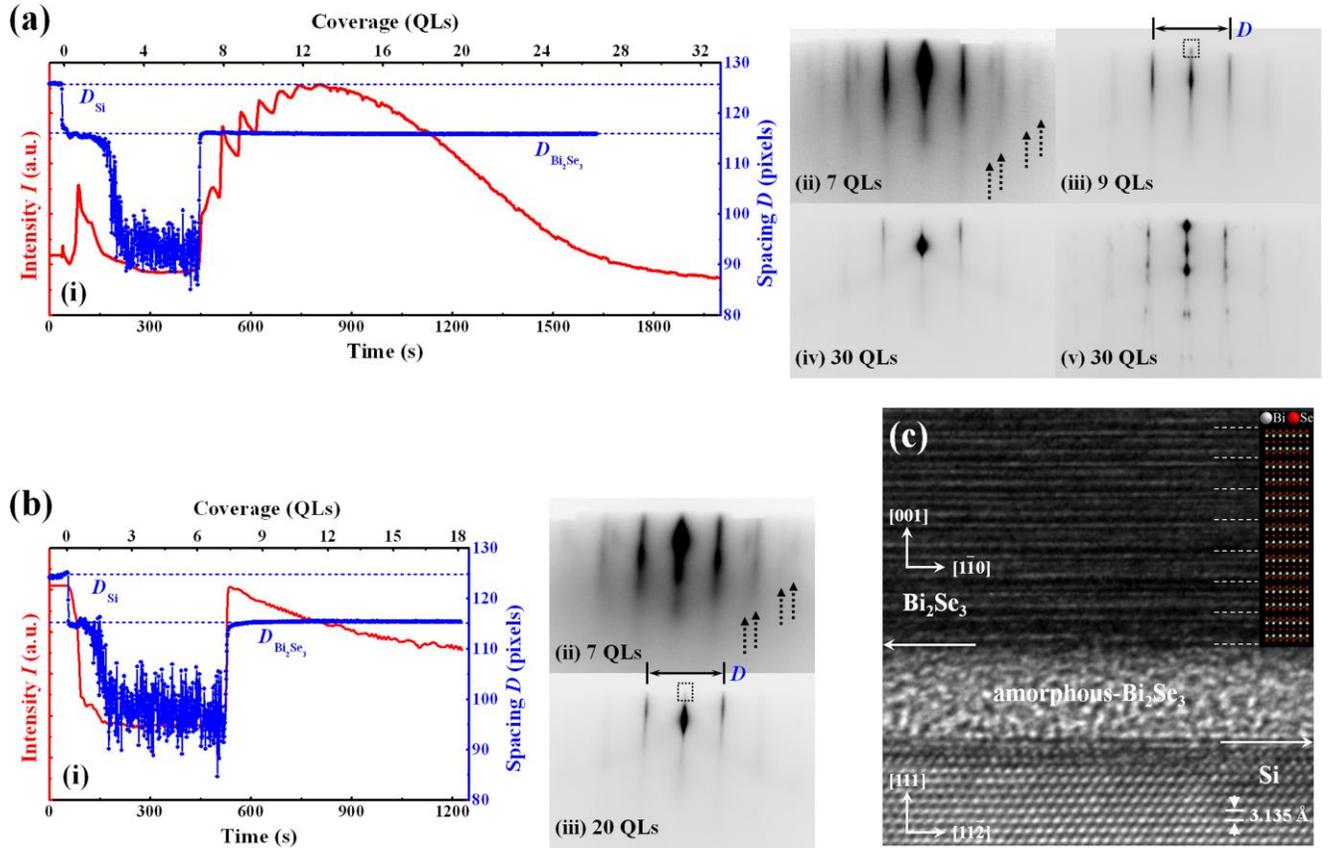

Figure 2

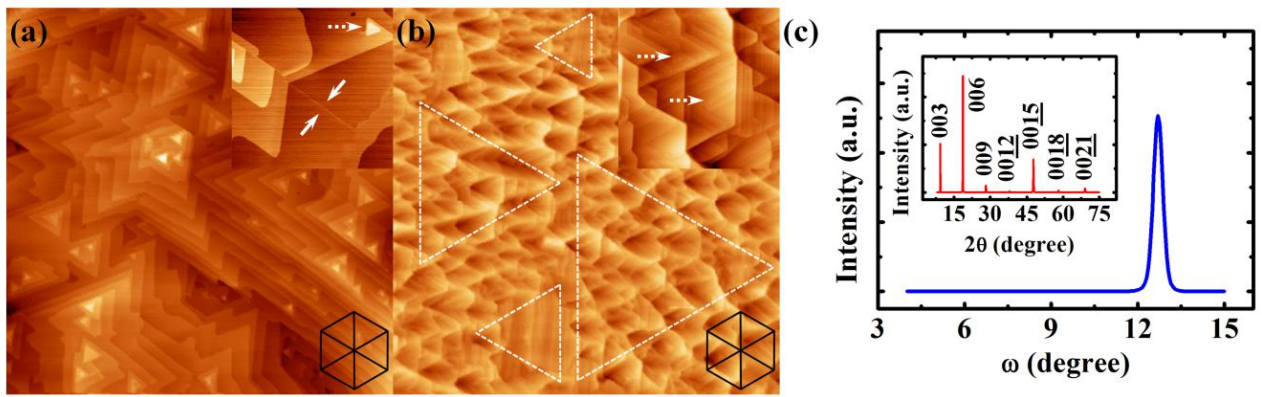

Figure 3

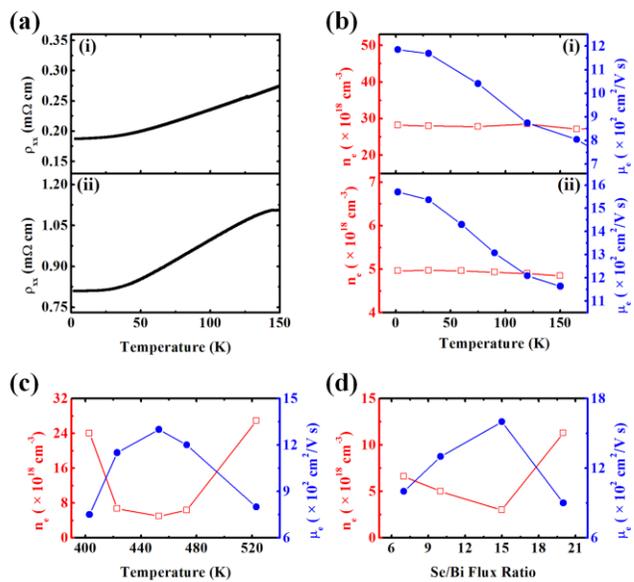

Figure 4

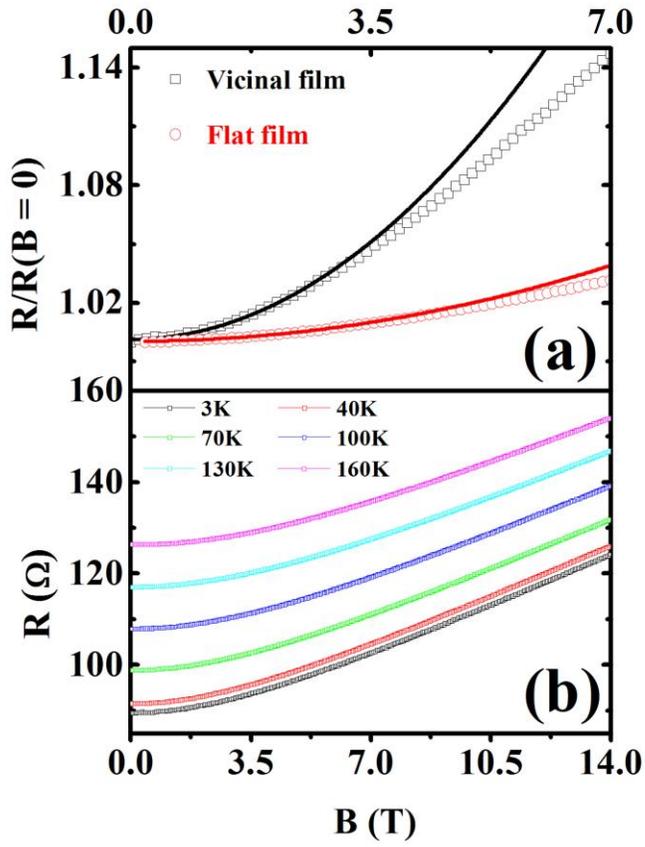